\documentclass[12pt,russian,english]{article}
\usepackage[T1]{fontenc}
\usepackage[cp1252]{inputenc}
\usepackage{textcomp}
\usepackage{amsmath}
\usepackage{amssymb}
\usepackage{esint}

\makeatletter

\DeclareRobustCommand{\cyrtext}{%
  \fontencoding{T2A}\selectfont\def\encodingdefault{T2A}}
\DeclareRobustCommand{\textcyr}[1]{\leavevmode{\cyrtext #1}}
\AtBeginDocument{\DeclareFontEncoding{T2A}{}{}}

\@ifundefined{date}{}{\date{}}

\newtheorem{theorem}{Theorem}

\newtheorem{corollary}{Corollary}

\newtheorem{definition}[theorem]{Definition}

\newtheorem{lemma}{Lemma}

\newtheorem{proposition}{Proposition}

\usepackage{babel}

\usepackage{babel}

\makeatother

\usepackage{babel}
\begin{document}

\title{Liouville ergodicity of linear multi-particle hamiltonian system
with one marked particle velocity flips}

\author{Lykov A. A., Malyshev V. A.%
\thanks{Faculty of Mechanics and Mathematics, Lomonosov Moscow State University.
Vorobievy Gory, Main Building, 119991, Moscow Russia, malyshev2@yahoo.com%
}}
\maketitle
\begin{abstract}
We consider multi-particle systems with linear deterministic hamiltonian
dynamics. Besides Liouville measure it has continuum of invariant
tori and thus continuum of invariant measures. But if one specified
particle is subjected to a simple linear deterministic transformation
(velocity flip) in random time moments, we prove convergence to Liouville
measure for any initial state. For the proof it appeared necessary
to study non-linear transformations on the energy surface. 
\end{abstract}
\tableofcontents

\pagebreak

\section{Introduction}

Ergodicity problem for hamiltonian multi-particle systems produced
many deep results. First of all, many examples of non-ergodic systems
appeared - linear, non-linear with additional integrals and close
to them (KAM theory). One could expect then that for generic hamiltonians
one also has non-ergodic behaviour. However, as far as we know, this
is still an open difficult problem with many partial results, see
\cite{simanyi_1,villani_1}, and, after the century history of the
ergodicity hypothesis it is reasonable to look for simpler alternative
approaches to it.

Namely, one could assume that any physical system has always some
contact with external world. Such contact can be of quite various
extent: 1) all particles can have contact with external world and
stochastic behaviour (for example, with dynamics of Glauber type),
2) only particles on the boundary, etc. But then it is quite natural
to ask - what is the minimal contact which definitely provides ergodic
behavior. Possible reformulation of the ergodicity hypothesis could
be the following: for generic system even the minimalistic contact
produces ergodic behaviour.

Here we consider an example of such minimalistic contact which consists,
first of all, in that we allow some contact with external world for
only one (marked) particle. In our earlier papers \cite{LM_1,LM_4}
this particle was subjected to some random force, that garantied convergence
to Gibbs equilibrium. Here we assume even less randomness. Namely,
the marked particle is subjected to a simple deterministic transformation
(velocity flip, very popular in other problems \cite{bernardin_olla,lukkarinen_2014,Simon,vf_1994,vp_2011,vp_2012})
but in discrete random time moments 
\begin{equation}
0<t_{1}<\ldots<t_{m}<\ldots\label{time_sequence}
\end{equation}
Liouville measure is evidently invariant w.r.t. such dynamics on the
energy surface. We prove then that ergodicity in the stronger form
holds - for any initial state we have convergence to Liouville measure
on the energy surface.

It is very interesting that it works even for the systems having the
worst possible non-ergodic behaviour - linear systems. The only price
we pay is that ergodicity holds not for any linear system but for
almost any - this is purely algebraic phenomenon which was discussed
in our earlier papers and cannot be avoided. It seems reasonable that
the same result holds for non-linear systems as well (following the
common belief that the latter have better mixing properties than linear
systems). 

The paper is naturally subdivided in two parts. The first part uses
no probability at all but only elaborate non-linear analysis to prove
that the trajectory visits all invariant tori and even any point -
we had to use coordinates on the energy surface where velocity flips
are strongly non-linear. Second part, on the contrary, essentially
uses non-trivial parts of Markov processes theory on continuous state
space. 

Now we come to rigorous definitions.

\section{Model and Main Results}

\paragraph{Hamiltonian dynamics}

We consider the linear space 
\[
L=\mathbb{R}^{2N}=\{\psi=\left(\begin{array}{c}
q\\
p
\end{array}\right):\ q=(q_{1},...,q_{N})^{T},\ p=(p_{1},...,p_{N})^{T},\ q_{i},p_{i}\in\mathbb{R}\},
\]
where $T$ denotes transposition (thus $\psi$ is a column-vector).
It can be presented as the direct sum $L=l_{N}^{(q)}\oplus l_{N}^{(p)}$
of two orthogonal (coordinate and momenta) spaces of dimension $N$
with the standard scalar product in $\mathbb{R}^{2N}$ 
\[
(\psi,\psi')_{2}=(q,q')_{2}+(p,p')_{2}=\sum_{i=1}^{N}(q_{i}q_{i}'+p_{i}p_{i}')
\]
We consider quadratic hamiltonian 
\begin{equation}
H(\psi)=\sum_{k=1}^{N}\frac{p_{k}^{2}}{2}+U(q),U(q)=\frac{1}{2}(q,Vq)_{2}\label{Ham1}
\end{equation}
where the matrix $V>0$ acting in $\mathbb{R}^{N}$ is assumed to
be real and positive definite (thus the particles cannot escape to
infinity). This defines hamiltonian system of linear ODE with $k=1,\ldots,N$
\begin{equation}
\dot{q}_{k}=p_{k},\dot{p}_{k}=-\sum_{l=1}^{N}V_{kl}q_{l},\label{L1}
\end{equation}
For any $h>0$ define the constant energy surface 
\[
\mathcal{M}_{h}=\{\psi\in L:H(\psi)=h\}
\]
Then $\mathcal{M}_{h}$ is a smooth manifold (ellipsoid) in $L$ of
codimension $1$.

\paragraph{Allowed hamiltonians}

Define the mixing subspace 
\[
L_{-}=L_{-}(V)=\{\left(\begin{array}{c}
q\\
p
\end{array}\right)\in L:q,p\in l_{V}\}
\]
where $l_{V}=l_{V,1}$ is the subspace of $\mathbb{R}^{N}$, generated
by the vectors $V^{k}e_{1},\ k=0,1,2\ldots$, where $e_{1},...,e_{N}$
is the standard basis in $\mathbb{R}^{N}$.

Let $\mathbf{V}$ be the set of all positive-definite $(N\times N)$-matrices,
and let $\mathbf{V}^{+}\subset\mathbf{V}$ be the subset of matrices
for which

\begin{equation}
L_{-}(V)=L\label{L201406141}
\end{equation}
The set of $V\in\mathbf{V}$ such that their eigenvalues, denoted
by $\omega_{1}^{2},\ldots,\omega_{N}^{2}$, are independent over the
field of rational numbers, is denoted by $\mathbf{V}_{ind}$.

\begin{lemma}\label{V-sets}

The set $\mathbf{V}^{+}$ is open and everywhere dense (assuming topology
of $R^{\frac{N(N+1)}{2}}$) in $\mathbf{V},$ and the set $\mathbf{V}^{+}\cap\mathbf{V}_{ind}$
is dense both in $\mathbf{V}^{+}$and in $\mathbf{V}.$

\end{lemma}

See more in section 5.3.

\paragraph{Piecewise deterministic process}

Assume that at time moments (\ref{time_sequence}) the following deterministic
transformation $I:L\to L$ occurs: all $q_{k},p_{k}$ are left unchanged,
except for $p_{1}$, the sign of which becomes inverted 
\[
p_{1}(t_{m}-0)\to p_{1}(t_{m})=-p_{1}(t_{m}-0),m\geq1
\]
For example, one can consider $L$ as the phase space for $N$ identical
point particles in $\mathbb{R}$, with mass $m=1$, and real numbers
$q_{i},p_{i}$ are their coordinates and velocities (momenta). Then
this transformation can be interpreted as the elastic collision of
the particle $1$ with a wall. Alternatively, taking $dN$ instead
of $N$, one can imagine $N$ particles in $R^{d}$ where only one
velocity component of particle 1 is flipped. Reflections w.r.t. any
hyperplane in $R^{d}$ could be considered quite similarly.

In-between these moments the system evolves via hamiltonian dynamics
(\ref{L1}).

With $(2N\times2N)$-matrix 
\[
A=\left(\begin{array}{cc}
0 & E\\
-V & 0
\end{array}\right)
\]
the system (\ref{L1}) can be rewritten as 
\begin{equation}
\dot{\psi}=A\psi.\label{LHamVec}
\end{equation}
and the solution $\psi(t)$ of (\ref{LHamVec}) with initial vector
$\psi(0)$ will be 
\[
\psi(t)=e^{tA}\psi(0).
\]
For given sequence (\ref{time_sequence}) the dynamics of our process
is defined for $t_{m}\leq t<t_{m+1}$ as 
\[
\psi(t)=e^{A(t-t_{m})}Ie^{A\tau_{m}}Ie^{A\tau_{m-1}}I...Ie^{A\tau_{2}}Ie^{A\tau_{1}}\psi(0)
\]
where $\tau_{1}=t_{1},\tau_{2}=t_{2}-t_{1},...,\tau_{m}=t_{m}-t_{m-1},...$.
For any $t\geqslant0$ define linear maps $L\to L$ 
\[
J(t)\psi=Ie^{tA}\psi,\ \psi\in L
\]
It is clear that $\mathcal{M}_{h}$ is invariant w.r.t. $J(t)$ for
any $h>0$ and $t>0$. For any $\psi\in L$ and any integer $m\geqslant1$
define the set of states 
\[
\mathcal{J}_{m}(\psi)=\{J(\tau_{m})\ldots J(\tau_{1})\psi:\ 0\leq\tau_{1},...,\tau_{m}\}\subset\mathcal{M}_{h}
\]
which the system can visit at the $m$-th flip.

\begin{theorem}[covering theorem] \label{th_cover} Assume that $V\in\mathbf{V}^{+}\cap\mathbf{V}_{ind}$,
then there exists $m\geqslant1$ such that for any $\psi\in L$ we
have 
\[
\mathcal{J}_{m}(\psi)=\mathcal{M}_{h}
\]
\end{theorem}

We introduce randomness by the following assumption 

\textbf{A0}): positive random variables
\[
\tau_{1}=t_{1},\tau_{2}=t_{2}-t_{1},...,\tau_{n}=t_{n}-t_{n-1},...
\]
are assumed to be independent, identically distributed with measure
$P_{\tau}=\rho(s)ds$ where the density $\rho$ (w.r.t. Lebesgue measure
$ds$) is positive everywhere on $R_{+}$, and moreover the first
moment $E\tau_{1}<\infty$. 

If, for example, $\tau_{i}$ have exponential distribution with the
density $\lambda\exp(-\lambda\tau),\ \lambda>0$, then it defines
Markov process $\psi(t)$ with right continuous deterministic trajectories
and random jumps. Such processes are often called piecewise deterministic
Markov processes, see for example \cite{piece-wise-det}. At the same
time, this can be considered as an example from random perturbation
theory, see \cite{Kifer} where the problem of invariant measures
is studied.

Let $\pi$ be Liouville measure on the energy surface, defined by
the surface form $d\sigma$ divided by $|\nabla H|$. It is well known
that $\pi$ is invariant w.r.t. hamiltonian dynamics and also w.r.t.
velocity flips. 

\begin{theorem}[convergence theorem]\label{th_erg} Assume that $V\in\mathbf{V}^{+}\cap\mathbf{V}_{ind}$.
Then under assumption $A0)$ for any initial $\psi(0)$ and any bounded
measurable real function $f$ on $\mathcal{M}_{h}$ we have a.s. 
\[
M_{f}(T)=^{def}\frac{1}{T}\int_{0}^{T}f(\psi(t))dt\to\pi(f)=^{def}\int_{\mathcal{M}_{h}}fd\pi
\]

\end{theorem} We call this property Liouville ergodicity.

\section{Deterministic part - proof of covering theorem}

\paragraph{Plan of the proofs}

In all assertions below we always assume that $V\in\mathbf{V}^{+}\cap\mathbf{V}_{ind}$.
Theorem \ref{th_erg} will follow from theorem \ref{th_cover}. Theorem
\ref{th_cover} will follow from the following weaker results.

\begin{theorem}[closure theorem] \label{th_closure} There exists
$m\geqslant1$ such that for all $\psi\in L$ we have 
\[
\overline{\mathcal{J}_{m}(\psi)}=\mathcal{M}_{h}
\]
\end{theorem}

The covering and closure theorems have simpler local analogue. And
moreover it shows that the dimension of $\mathcal{J}_{m}(\psi)$ grows
as $m$ for $m=1,2,...,2N-1$. For exact formulation we need some
definitions. For any $\tau_{1},...,\tau_{m}>0$ denote $J(\tau_{1},\ldots,\tau_{m})=J(\tau_{m})\ldots J(\tau_{1})$.
It can be considered as the mapping from $L$ to $\mathcal{M}_{h}$
for fixed $(\tau_{1},\ldots,\tau_{m})$, but also $J(\tau_{1},\ldots,\tau_{m})\psi$
can be considered as the map 
\[
J_{m}^{\psi}:\Omega_{m}\rightarrow\mathcal{M}_{h},\quad\bar{\tau}=(\tau_{1},\ldots,\tau_{m})\mapsto J(\tau_{1},\ldots,\tau_{m})\psi.
\]
from the $m$-dimensional orthant $\Omega_{m}=\{(\tau_{1},\ldots,\tau_{m}):\ \tau_{i}>0,\ \ i=1,\ldots,m\}=\mathbb{R}_{+}^{m},$
to $\mathcal{M}_{h},$ for fixed $\psi\in\mathcal{M}_{h}$.

\begin{theorem}[local covering theorem]\label{local_embed} For any
point $\psi\in\mathcal{M}_{h}$ and any $k=1,\ldots,2N-1$ the dimension
of $J_{k}^{\psi}\Omega_{k}$ is equal to $k$. Moreover, there exists
open subset $U\subset\Omega_{k}$ such that the mapping $J_{k}^{\psi}:U\hookrightarrow\mathcal{M}_{h}$
is a smooth embedding \end{theorem}

\subsection{Key definitions, notation and some intuition}

The sequence (\ref{time_sequence}) completely defines the trajectory,
or the path. Proof of convergence and covering theorems could be obtained,
in some sense, by ``summation'' over all possible paths. For this
we have to use various coordinate systems on the energy surface. To
get some intuition it is useful to see how it works for easier cases
$N=1,2$.

\paragraph{Action-angle coordinates}

Let $v_{1},\ldots,v_{N}$ be the orthonormal eigenvectors of $V$
in $\mathbb{R}^{N}$, and let $\omega_{1}^{2},\ldots,\omega_{N}^{2}$
be the corresponding eigenvalues of $V$. For any $k=1,\ldots,N$
define $2N$-vectors in $L$: 
\begin{equation}
Q_{k}=(v_{k},0)^{T},\quad P_{k}=(0,v_{k})^{T}\label{P_k_Q_k}
\end{equation}
The coordinates of $\psi\in L$ in the basis $Q_{1},P_{1},\ldots,Q_{N},P_{N}$
we denote by $(\tilde{q_{k}},\tilde{p_{k}})^{T}$, that is 
\begin{equation}
\psi=\sum_{k=1}^{N}\tilde{q}_{k}Q_{k}+\sum_{k=1}^{N}\tilde{p}_{k}P_{k}\label{P_k_Q_k_1}
\end{equation}
They correspond to coordinates and momenta of the independent <<quasiparticles>>,
that is one-dimensional oscillators, having energies $\frac{r_{k}^{2}}{2}$
where 
\[
r_{k}^{2}(\psi)=\tilde{p}_{k}^{2}+\omega_{k}^{2}\tilde{q}_{k}^{2}=(\psi,P_{k})_{2}^{2}+\omega_{k}^{2}(\psi,Q_{k})_{2}^{2}=(p,P_{k})_{2}^{2}+\omega_{k}^{2}(q,Q_{k})_{2}^{2},k=1,\ldots,N
\]
are the action coordinates of the point $\psi=(q,p)^{T}$. We agree
that $r_{k}(\psi)=\sqrt{r_{k}^{2}(\psi)}\geqslant0$. It is easy to
see that $r_{k}$ are integrals of the hamiltonian dynamics, that
is for any $t\geqslant0$ 
\begin{equation}
r_{k}^{2}(\psi)=r_{k}^{2}(e^{tA}\psi)\label{firstInt}
\end{equation}
The angle variables then are the angles for these oscillators.

The following assertions easily follow from the known facts, see for
example \cite{Arnold}, pp. 103, 272. But the proof is very elementary
and we give it for the reader's convenince in Appendix.

\begin{lemma}\label{L:action-angle}

Consider hamiltonian dynamics (\ref{L1}), then 
\begin{enumerate}
\item For any $r_{1}\geqslant0,\ldots,r_{N}\geqslant0$ the set 
\[
T(r_{1},\ldots,r_{N})=\{\psi\in L:r_{k}(\psi)=r_{k},k=1,...,N\}
\]
is invariant and diffeomorphic to torus of dimension $N-n$, where
$n$ equals the number of zeros among $r_{1},\ldots,r_{N}$. 
\item For any point $\psi\in L$ the closure of its orbit coincides with
the torus defined by it, that is 
\[
\overline{\{e^{tA}\psi:t\geqslant0\}}=T(r_{1}(\psi),\ldots,r_{N}(\psi)),
\]

\item Thus the torus defines the vector $\bar{r}=r_{1},r_{2},\ldots,r_{N}$.
Vice-versa, any such vector with non-negative coordinates uniquely
defines the torus. This torus lies on the energy surface $\mathcal{M}_{h}$
iff 
\[
\sum_{k=1}^{N}r_{k}^{2}=2h
\]

\end{enumerate}
\end{lemma}

For convenience we put $h=\frac{1}{2}$, and denote the set of all
invariant tori on $\mathcal{M}=\mathcal{M}_{\frac{1}{2}}$ by 
\[
\mathbf{T}=\{\bar{r}=(r_{1},\ldots,r_{N}):r_{i}\geqslant0,\sum_{k=1}^{N}r_{k}^{2}=1\}
\]

\paragraph{Case $N=1$}

Here $\omega_{1}^{2}$ can be arbitrary, and we take $h=\omega_{1}^{2}=1$,
let $S$ be the corresponding circle. Then the particle moves along
$S$ with constant angle velocity in the clock-wise direction. Covering
theorem is evident in this case (with $m=1$). However, for $N=1$
we have much stronger statement.

For any $\psi,\psi'$ denote $T(\psi,\psi')$ the minimal $t>0$ such
that $\psi'=e^{At}\psi$. Then $T_{0}=T(\psi,\psi)$ is the first
return time, or the time of complete rotation around the circle, and
for any $\psi$ 
\[
T(\psi,-\psi)=\frac{1}{2}T_{0}
\]

\begin{proposition}\label{N1}

For any $\psi,\psi'$ and any $t\geq T_{0}$ there exists $0<t_{1}=t_{1}(\psi,\psi',t)<T_{0}$
such that 
\[
e^{A(t-t_{1})}Ie^{At_{1}}\psi=\psi'
\]

\end{proposition}

For the proofs see Appendix.

\paragraph{$(r,\tilde{p})$-coordinates and $(r,p)$-coordinates }

Note that $I$ can be written as 
\[
I=E-2P_{1},
\]
where $E$ is the identity matrix and $P_{1}$ is the orthogonal projector
on the vector $g_{1}=(0,e_{1})^{T}\in L$. The expansion 
\[
e_{1}=\sum_{k=1}^{N}\beta_{k}v_{k}
\]
of the vector $e_{1}\in R^{N}$ defines the numbers $\beta_{k}$.
The $p_{k}$ and $\tilde{p}_{k}$ coordinates of the vector $p$ in
$R^{N}$ are related by the formulas 
\[
p=\sum_{i=1}^{N}p_{i}e_{i}=\sum_{k=1}^{N}\tilde{p}_{k}v_{k}
\]
\begin{equation}
\tilde{p}_{k}=\sum_{i=1}^{N}p_{i}(e_{i},v_{k})_{2},\,\,\, p_{k}=\sum_{i=1}^{N}\tilde{p}_{i}(v_{i},e_{k})_{2}\label{p_p_tilde}
\end{equation}
Note also that $I$ acts only on $p_{k}$-coordinates and $\tilde{p}_{k}$-coordinates,
but not on $q_{k}$ and $\tilde{q_{k}}$. Thus the following notation
is justified 
\begin{equation}
Ip=p-2(p,e_{1})_{2}e_{1}=\sum_{k=1}^{N}\tilde{p}'_{k}v_{k},\,\,\,\tilde{p}'_{k}=\tilde{p}_{k}-2p_{1}\beta_{k}\label{I_on_tilda}
\end{equation}
where 
\[
p_{1}=(p,e_{1})_{2}=\sum_{k=1}^{N}\tilde{p}_{k}\beta_{k}.
\]
By (\ref{p_p_tilde}) the velocity flip changes the action variables
as follows 
\[
r_{k}^{2}(I\psi)=(Ip,v_{k})_{2}^{2}+\omega_{k}^{2}(q,v_{k})_{2}^{2}=(\tilde{p}'_{k})^{2}+\omega_{k}^{2}(q,v_{k})_{2}^{2}=\tilde{p}_{k}^{2}+4p_{1}^{2}\beta_{k}^{2}-4\tilde{p}_{k}p_{1}\beta_{k}+\omega_{k}^{2}(q,v_{k})_{2}^{2}=
\]
\[
=r_{k}^{2}(\psi)+4p_{1}^{2}\beta_{k}^{2}-4\tilde{p}_{k}p_{1}\beta_{k}=r{}_{k}^{2}(\psi)+4p_{1}^{2}\beta_{k}^{2}-4p_{1}\beta_{k}(p,v_{k})_{2}
\]

The following formula defines the time evolution of the tori, in terms
of initial action variable and momenta at time $t-0$. Namely, for
any $k$ 
\begin{equation}
r_{k}^{2}(Ie^{tA}\psi)=r_{k}^{2}(I\psi(t-0))=r_{k}^{2}(\psi)+4p_{1}^{2}(t-0)\beta_{k}^{2}-4p_{1}(t-0)\beta_{k}(p(t-0),v_{k})_{2},\label{r_k}
\end{equation}
where $\psi(t-0)=e^{tA}\psi=(q(t-0),p(t-0))^{T}\in L,\ p(t-0)=(p_{1}(t-0),\ldots,p_{N}(t-0))^{T}$.
We can rewrite (\ref{r_k}) as 
\begin{equation}
\bar{r}(J(t)\psi)=\Psi(\bar{r}(\psi),p(t-0)),\label{TorusTransform}
\end{equation}
where 
\[
\Psi=(\Psi_{1},\ldots,\Psi_{N}),\Psi_{k}^{2}(\bar{r},p)=r_{k}^{2}+4p_{1}^{2}\beta_{k}^{2}-4p_{1}\beta_{k}(p,v_{k})_{2}
\]
has the domain of definition 
\[
D(\Psi)=\{(\bar{r},p):\bar{r}\in\mathbf{T},p\in T_{p}(\bar{r})\}\subset\mathbf{T}\times\mathbb{R}^{N}
\]
where $T_{p}(\bar{r})$ is the projection of the torus $T(\bar{r})$
onto the space of momenta, that is the set of all $p\in\mathbb{R}^{N}$
such that for some $q\in\mathbb{R}^{N}$ we have $(q,p)^{T}\in T(\bar{r})$.
For any $\psi=\psi(0)=(q,p)^{T},q=q(0),p=p(0)$ consider the following
set of tori 
\[
\mathcal{T}_{1}(\psi)=\overline{\{\bar{r}(J(t)\psi):t\geqslant0\}}=\overline{\{\Psi(\bar{r}(\psi),p(t-0)):t>0,p(0)=p\}}\subset\mathbf{T}.
\]

From continuity of $\Psi$ it follows that $\mathcal{T}_{1}(\psi)$
depends only on the invariant torus containing $\psi$, that is if
$\bar{r}(\psi)=\bar{r}(\psi')$ for two points $\psi,\psi'\in L$,
then 
\begin{equation}
\mathcal{T}_{1}(\psi)=\mathcal{T}_{1}(\psi')\label{only_from_torus_1}
\end{equation}
Then, the following notation is correct 
\begin{equation}
\mathcal{T}_{1}(\psi)=\mathcal{T}_{1}(\bar{r}(\psi))\label{only_from_torus_2}
\end{equation}
It is useful to note that by continuity of $\Psi$ for any $\bar{r}\in\mathbf{T}$
we have 
\[
\mathcal{T}_{1}(\bar{r})=\{\Psi(\bar{r},p):p\in T_{p}(\bar{r})\}
\]

\paragraph{Case $N=2$}

Here we will only give intuitive arguments why for any initial state
$\psi$, in finite number of jumps, we can enter any torus. From (\ref{r_k})
one can get the following formula for the evolution of the action
variables 
\[
r_{1}^{2}(J(t)\psi)-r_{1}^{2}(\psi)=4\beta_{1}\beta_{2}(\tilde{p}_{1}(t-0)\beta_{1}+\tilde{p}_{2}(t-0)\beta_{2})(\tilde{p_{2}}(t-0)\beta_{1}-\tilde{p_{1}}(t-0)\beta_{2}),
\]
and also 
\[
r_{2}^{2}(J(t)\psi)=1-r_{1}^{2}(J(t)\psi)
\]
Denote the right-hand side of the first formula by $D(t)$. It appears
that for some $t'$ the set $\{D(t):0\leq t\leq t'\}$ contains an
open interval, and moreover its length has lower bound $\delta>0$
uniform in $\psi$. Thus the measure of the set of visited tori, after
each application of $I$ enlarges by additive constant. It follows
that all tori may be visited for finite number of $I$ transformations.
To see this for general $N$ is more difficult.

\subsection{Proof of closure theorem}

\subsubsection{Contraction property}

Define the function $\rho$ in $\mathbf{T}$ 
\[
\rho(\bar{r},\bar{r}')=\sum_{k=1}^{N}|r_{k}^{2}-(r'_{k})^{2}|,
\]
where $\bar{r}=(r_{1},\ldots,r_{N}),\ \bar{r}'=(r'_{1},\ldots,r'_{N})$.
Denote $\bar{r}^{*}=(|\beta_{1}|,\ldots,|\beta_{N}|)\in\mathbf{T}$,
thus it is the invariant torus, containing the point $g_{1}=\sqrt{2}(0,e_{1})^{T}\in L$.
Note that all $\beta_{k}$ are nonzero, that follows from the assumption
$V\in\mathbf{V}^{+}$, see section 7.3. The point $g_{1}$ corresponds
to the configuration of particles where all particle have zero velocity
and zero coordinates, except for the particle $1$.

\begin{theorem}[contraction theorem] \label{contractTheorem} For
any $\bar{r}=(r_{1},\ldots,r_{N})\in\mathbf{T}$ we have the following
contraction bound 
\[
\rho(\mathcal{T}_{1}(\bar{r}),\bar{r}^{*})\leqslant(1-c(\bar{r}))\rho(\bar{r},\bar{r}^{*}),
\]
where the constant $c(\bar{r})$ is given by 
\begin{equation}
c(\bar{r})=\frac{1}{\max\{1,D^{2}(\bar{r})\}},\quad D(\bar{r})=\max_{k=1,\ldots,N}\frac{r_{k}}{|\beta_{k}|}-\min_{k=1,\ldots,N}\frac{r_{k}}{|\beta_{k}|}\label{ceq}
\end{equation}
\end{theorem}

Proof. We will find point $p^{'}\in T_{p}(\bar{r})$ such that 
\[
\rho(\Psi(\bar{r},p^{'}),\bar{r}^{*})=(1-c(\bar{r}))\rho(\bar{r},\bar{r}^{*}).
\]
Note first that in coordinates $\tilde{p}$ the set $T_{p}(\bar{r})$
is the $N$-dimensional cube with sides $(2r_{1},\ldots,2r{}_{N})$,
this follows from the oscillator representation. In other words, the
point $p=(\tilde{p}_{1},\ldots,\tilde{p}_{N})^{T}\in T_{p}(\bar{r})$
iff for any $k=1,\ldots,N$ 
\begin{equation}
|\tilde{p_{k}}|\leqslant r_{k},\label{pteq}
\end{equation}
For $k=1,\ldots,N$ denote 
\[
\gamma_{k}=\frac{r_{k}}{|\beta_{k}|}.
\]
and denote the minimal of them by $\gamma_{n}$ and maximal by $\gamma_{N}$.
We will need the following functions 
\[
f^{\pm}(x)=\frac{1}{2}\left(x\pm\sqrt{x^{2}+c(1-x^{2})}\right)
\]
of $x\in\mathbb{R}$, where the constant $c$ is defined in the formula
(\ref{ceq}). As $c\leqslant1$, then for all $x$ 
\[
x^{2}+c(1-x^{2})\geq0
\]

Define the point $p^{'}\in\mathbb{R}^{N}$ with coordinates $(\tilde{p}_{1}^{'},\ldots,\tilde{p}_{N}^{'})$
in the basis $v_{1},\ldots,v_{N}$ by the formula 
\[
\tilde{p}_{k}^{'}=y\beta_{k}-c\frac{\beta_{k}^{2}-r_{k}^{2}}{4y\beta_{k}},\quad y=f^{+}(\gamma_{n}).
\]
and find the value of $\Psi(\bar{r},p^{'})$. As 
\[
p_{1}=(p^{'},e_{1})_{2}=\sum_{k=1}^{N}\beta_{k}\tilde{p}_{k}^{'}=y\sum_{k=1}^{N}\beta_{k}^{2}-\frac{c}{4y}\sum_{k=1}^{N}(\beta_{k}^{2}-r_{k}^{2})=y.
\]
then for any $k=1,\ldots,N$ we have 
\[
\Psi_{k}^{2}=r_{k}^{2}+4p_{1}^{2}\beta_{k}^{2}-4p_{1}\beta_{k}\tilde{p}_{k}^{'}=r_{k}^{2}+4y^{2}\beta_{k}^{2}-4y\beta_{k}\left(y\beta_{k}-c\frac{\beta_{k}^{2}-r_{k}^{2}}{4y\beta_{k}}\right)=r_{k}^{2}+c(\beta_{k}^{2}-r_{k}^{2}).
\]
Then we have the distance 
\[
\rho(\Psi(\bar{r},p^{'}),\bar{r}^{*})=\sum_{k=1}^{N}|\Psi_{k}^{2}-\beta_{k}^{2}|=(1-c)\rho(\bar{r},\bar{r}^{*}).
\]
\begin{lemma}\label{lemma_cube} $p^{'}\in T_{p}(\bar{r})$ that
is the inequalities (\ref{pteq}) hold.

\end{lemma}

Proof consists of simple calculations and is given in Appendix.

From the contraction theorem only, one cannot prove the closure theorem
because $c=c(\bar{r})$ depends on $\bar{r}$. That is why we need
one more assertion. For any $\bar{r}=(r_{1},\ldots,r_{N})\in\mathbf{T}$
put 
\[
A(\bar{r})=\min_{k=1,\ldots,N}\frac{r_{k}}{|\beta_{k}|},\quad B(\bar{r})=\max_{k=1,\ldots,N}\frac{r_{k}}{|\beta_{k}|},\quad\Delta(\bar{r})=B^{2}(\bar{r})-A^{2}(\bar{r}).
\]
The function $\Delta(\bar{r})$ defined on $\mathbf{T}$ also looks
like a distance to the point $\bar{r}^{*}$. Let us explain this in
more detail. In particular, $\Delta(\bar{r})=0$ iff $r_{k}=|\beta_{k}|$
for all $k=1,\ldots,N$. Moreover, as 
\begin{equation}
\sum_{k=1}^{N}r_{k}^{2}=1\label{sumRk}
\end{equation}
then, if $\Delta(\bar{r})$ is small then the numbers $r_{k}$ are
close to $|\beta_{k}|$. Even more, it gives an upper bound for the
distance, more exactly for any $\bar{r}\in\mathbf{T}$ 
\begin{equation}
\rho(\bar{r},\bar{r}^{*})=\sum_{k}|r_{k}^{2}-\beta_{k}^{2}|=\sum_{k}\beta_{k}^{2}|\frac{r_{k}^{2}}{\beta_{k}^{2}}-1|\leqslant\sum_{k}\beta_{k}^{2}(B^{2}(\bar{r})-A^{2}(\bar{r}))=\sum_{k}\beta_{k}^{2}\Delta(\bar{r})\leqslant\Delta(\bar{r})\label{delta}
\end{equation}
The following result shows that each velocity flip at appropriate
moments makes point $g_{1}$ closer by 1.

\begin{corollary} \label{lemma_corrolary_contract} For any point
$\bar{r}\in\mathbf{T}$ there exists $p\in T_{p}(\bar{r})$ such that
\[
\Delta(\Psi(\bar{r},p))\leqslant\max\{\Delta(\bar{r})-1,0\}
\]
\end{corollary}

Proof. From the proof of contraction theorem it follows that there
exists $p^{'}\in T_{p}(\bar{r})$ such that 
\[
\Psi_{k}^{2}(\bar{r},p^{'})=r_{k}^{2}+c(\bar{r})(\beta_{k}^{2}-r_{k}^{2}).
\]
Denote $\bar{r}'=(r'_{1},\ldots,r'_{N})=\Psi(\bar{r},p^{'})$. Without
loss of generality one can choose the indices so that 
\[
\frac{r'_{1}}{|\beta_{1}|}=A(\bar{r}'),\quad\frac{r'_{N}}{|\beta_{N}|}=B(\bar{r}').
\]
Then 
\[
\Delta(\bar{r}')=\frac{(r'_{N})^{2}}{\beta_{N}^{2}}-\frac{(r'_{1})^{2}}{\beta_{1}^{2}}=(1-c(\bar{r}))\left(\frac{r_{N}^{2}}{\beta_{N}^{2}}-\frac{r_{1}^{2}}{\beta_{1}^{2}}\right).
\]
from where we get 
\begin{equation}
\Delta(\bar{r}')\leqslant(1-c(\bar{r}))\Delta(\bar{r}).\label{DeltaEq}
\end{equation}

\[
c(\bar{r})=\frac{1}{\max\{1,(B(\bar{r})-A(\bar{r}))^{2}\}}.
\]
Note that the following inequality holds 
\[
(B-A)^{2}\leqslant B^{2}-A^{2}=\Delta.
\]
and then 
\[
c(\bar{r})=\frac{1}{\max\{1,(B(\bar{r})-A(\bar{r}))^{2}\}}\geqslant\frac{1}{\max\{1,B^{2}(\bar{r})-A^{2}(\bar{r})\}}=\frac{1}{\max\{1,\Delta(\bar{r}\}}.
\]
From this inequality and the bound (\ref{DeltaEq}) finally we have
\[
\Delta(\bar{r}')\leqslant\left(1-\frac{1}{\max\{1,\Delta(\bar{r})\}}\right)\Delta(\bar{r})=\max\{\Delta(\bar{r})-1,0\}.
\]
$\blacktriangledown$

\subsubsection{Closure theorem}

Before proving it we have to prove even weaker assertion.

\begin{lemma} \label{lemma_g_1-1} There exists integer $m\geqslant1$
such that for any $\psi\in\mathcal{M}$ 
\[
g_{1}\in\overline{\mathcal{J}_{m}(\psi)}
\]

\end{lemma}

In other words there exists $m$ such that for any $\psi\in\mathcal{M}$
and any $\epsilon>0$ one can find moments $0<t_{1}<\ldots<t_{m}$
so that 
\[
||J(\tau_{m})\ldots J(\tau_{1})\psi-g_{1}||_{2}<\epsilon,
\]
By continuity of maps $\Psi$ and $\Delta$, and by Corollary \ref{lemma_corrolary_contract}
it follows that for any point $\psi\in L$ and any $\epsilon>0$ one
can find time moment $t\geqslant0$ such that 
\[
\Delta(\bar{r}(J(t)\psi))=\Delta(\Psi(\bar{r}(\psi),p(t-0)))\leqslant\max\{\Delta(\bar{r}(\psi))-1,0\}+\epsilon.
\]
It follows that there exists $m\geqslant1$ such that for any $\epsilon>0$
there exist time moments $t_{1},\ldots,t_{m}\geqslant0$ such that
\[
\Delta(\bar{r}(\psi'))\leqslant\epsilon,\quad\psi'=J(\tau_{m})\ldots J(\tau_{1})\psi.
\]
and moreover one can take $m=[\Delta(\bar{r}(\psi))]+1$. But as for
any $\bar{r}\in\mathbf{T}$ the following inequality holds 
\[
\Delta(\bar{r})\leqslant\max_{k=1,\ldots,N}\frac{1}{\beta_{k}^{2}},
\]
$m$ can be chosen uniformly in $\psi$. By formula (\ref{delta})
we get 
\[
\rho(\bar{r}(\psi'),\bar{r}^{*})\leqslant\epsilon.
\]
Alternatively it is evident that there exists constant $c>0$ such
that for any $\epsilon'>0$ there is $t\geqslant0$ such that 
\[
||e^{tA}\psi'-g_{1}||_{2}^{2}\leqslant c\rho(\bar{r}(\psi'),\bar{r}^{*})+\epsilon'.
\]
As $\epsilon$ and $\epsilon'$ are arbitrary and by $e^{tA}=J(0)J(t)$
we get the proof. $\blacktriangledown$

Now we prove the closure theorem. Define another norm on $L$ 
\[
||\psi||_{H}=\sqrt{H(\psi)}
\]
Let us fix two points $\psi,\psi'\in\mathcal{M}$ and show that $\psi'\in\overline{\mathcal{J}_{m}(\psi)}$
for some $m\geqslant1$ not depending on $\psi,\psi'$. By lemma \ref{lemma_g_1-1}
there exist $\tau_{1},...,\tau_{m}>0$ and $\tau'_{1},...,\tau'_{m}>0$
such that 
\[
||J(\tau_{1},\ldots,\tau_{m})\psi-J(\tau'_{1},\ldots,\tau'_{m})\psi'||_{H}<\epsilon,
\]
It is clear that the transform $J(\tau'_{1},\ldots,\tau'_{m})$ is
invertible and conserves the norm $||\ ||_{H}$, that is why 
\[
||J^{-1}(\tau'_{1},\ldots,\tau'_{m})J(\tau_{1},\ldots,\tau_{m})\psi-\psi'||_{H}<\epsilon.
\]
One can find $m'\leqslant2m$ and $\tau''_{1},...,\tau''_{m'}>0$
such that 
\begin{equation}
||J(\tau''_{1},\ldots,\tau''_{m'})J(\tau_{1},\ldots,\tau_{m})\psi-\psi'||_{H}<2\epsilon,\label{J_minus_1}
\end{equation}
In fact 
\begin{equation}
J^{-1}(\tau_{1},\ldots,\tau_{m})=e^{-\tau_{1}A}Ie^{-\tau_{2}A}I\ldots e^{-\tau_{m}A}I\label{J}
\end{equation}
As $e^{tA}$ conserves the norm $||.||_{H}$ and $I=J(0)$, it is
sufficient to show that for any point $\psi\in\mathcal{M}$, any $\epsilon>0$
and $t\geqslant0$ there exists $s=s(\psi,t,\epsilon)$ such that
\[
||e^{-tA}\psi-e^{sA}\psi||_{H}<\epsilon
\]
But as the closure of the orbit of any point is the torus, it is Poincare
recurrence theorem.

From (\ref{J_minus_1}) and equivalence of the norms $||\ ||_{2}$
and $||\ ||_{H}$ the closure theorem follows.

\subsection{Proof of local covering theorem}

Fix $\psi$. It is sufficient to show that for some point $\bar{\tau}\in\Omega_{k}$
the rank of $dJ_{k}^{\psi}(\bar{\tau})$ equals $k$. The columns
of the Jacobian matrix $J_{k}^{\psi}$ are the vectors 
\begin{align*}
\theta_{k}(\tau)=\frac{d}{d\tau_{k}}J(\tau_{1},\ldots,\tau_{k})\psi= & \ Ie^{\tau_{k}A}AJ(\tau_{1},\ldots,\tau_{k-1})\psi\\
\theta_{i}(\tau)=\frac{d}{d\tau_{i}}J(\tau_{1},\ldots,\tau_{k})\psi= & \ Ie^{\tau_{k}A}\frac{d}{d\tau_{i}}J(\tau_{1},\ldots,\tau_{k-1})\psi,\quad i=1,\ldots,k-1.
\end{align*}
Denote $\dim\langle w_{1},w_{2},...\rangle$ the dimension of the
linear span of vectors $w_{1},w_{2},...$. As the mapping $Ie^{\tau_{k}A}$
is non-degenerate, then for any $k=1,2,...$ one should derive that
there is $\bar{\tau}=(\tau_{1},\ldots,\tau_{k})\in\Omega_{k}$ such
that 
\[
\dim\langle\theta_{1}(\tau),\ldots,\theta_{k}(\tau)\rangle=\dim\langle AJ(\bar{\tau}_{c})\psi,\frac{d}{d\tau_{1}}J(\bar{\tau}_{c})\psi,\ldots,\frac{d}{d\tau_{k-1}}J(\bar{\tau}_{c})\psi\rangle=k,
\]
where $\bar{\tau}_{c}=(\tau_{1},\ldots,\tau_{k-1})\in\Omega_{k-1}$.
We shall do it by induction in $k$. Remind that for any $\psi\in L$
\[
I\psi=\psi-2(\psi,g_{1})_{2}g_{1}.
\]
In fact, for $k=1$ this is trivial.

Induction step. Assuming that we have proved the assertion for $k\leq2N-2$,
we shall prove it for $k+1$.

For $\bar{\tau}=(\tau_{1},\ldots,\tau_{k})\in\Omega_{k}$ denote $\bar{\tau}_{c}=(\tau_{1},\ldots,\tau_{k-1})\in\Omega_{k-1}$.
Then 
\begin{align*}
 & AJ(\bar{\tau})\psi=AIe^{\tau_{k}A}J(\tau_{1},\ldots,\tau_{k-1})\psi=Ae^{\tau_{k}A}J(\tau_{1},\ldots,\tau_{k-1})\psi+a_{0}(\bar{\tau})Ag_{1}=\\
 & =e^{\tau_{k}A}\left(AJ(\tau_{1},\ldots,\tau_{k-1})\psi+a_{0}(\bar{\tau})Ae^{-\tau_{k}A}g_{1}\right)=e^{\tau_{k}A}\left(AJ(\bar{\tau}_{c})\psi+a_{0}(\bar{\tau})Ae^{-\tau_{k}A}g_{1}\right),
\end{align*}
where $a_{0}(\bar{\tau})=-2(e^{\tau_{k}A}J(\bar{\tau}_{c})\psi,g_{1})_{2}$.
Similarly we get equalities for the derivatives for $i=1,\ldots,k-1$:
\begin{align*}
 & \frac{d}{d\tau_{i}}J(\bar{\tau})\psi=Ie^{\tau_{k}A}\frac{d}{d\tau_{i}}J(\tau_{1},\ldots,\tau_{k-1})\psi=e^{\tau_{k}A}\frac{d}{d\tau_{i}}J(\tau_{1},\ldots,\tau_{k-1})\psi+a_{i}(\bar{\tau})g_{1}=\\
 & =e^{\tau_{k}A}\left(\frac{d}{d\tau_{i}}J(\bar{\tau}_{c})\psi+a_{i}(\bar{\tau})e^{-\tau_{k}A}g_{1}\right),\quad a_{i}(\bar{\tau})=-2(e^{\tau_{k}A}\frac{d}{d\tau_{i}}J(\bar{\tau}_{c})\psi,g_{1})_{2}.
\end{align*}
Derivative in $\tau_{k}$ has the following expansion 
\begin{align*}
 & \frac{d}{d\tau_{k}}J(\bar{\tau})\psi=IAe^{\tau_{k}A}J(\tau_{1},\ldots,\tau_{k-1})\psi=Ae^{\tau_{k}A}J(\tau_{1},\ldots,\tau_{k-1})\psi+a_{k}(\bar{\tau})g_{1}=\\
 & =e^{\tau_{k}A}\left(AJ(\bar{\tau}_{c})\psi+a_{k}(\bar{\tau})e^{-\tau_{k}A}g_{1}\right),\quad a_{k}(\bar{\tau})=-2(Ae^{\tau_{k}A}J(\bar{\tau}_{c})\psi,g_{1})_{2}.
\end{align*}

Then by nondegeneracy of the mapping $e^{\tau_{k}}A$, we get that
the dimension of subspace 
\[
\langle AJ(\bar{\tau})\psi,\frac{d}{d\tau_{1}}J(\bar{\tau})\psi,\ldots,\frac{d}{d\tau_{k}}J(\bar{\tau})\psi\rangle
\]
equals the dimension $\aleph$ of the linear span of $(k+1)$ following
vectors 
\begin{align*}
 & AJ(\bar{\tau}_{c})\psi+a_{0}(\bar{\tau})Ae^{-\tau_{k}A}g_{1},AJ(\bar{\tau}_{c})\psi+a_{k}(\bar{\tau})e^{-\tau_{k}A}g_{1},\\
 & \frac{d}{d\tau_{1}}J(\bar{\tau}_{c})\psi+a_{1}(\bar{\tau})e^{-\tau_{k}A}g_{1},\ldots,\frac{d}{d\tau_{k-1}}J(\bar{\tau}_{c})\psi+a_{k-1}(\bar{\tau})e^{-\tau_{k}A}g_{1}.
\end{align*}
By inductive assumption there exists point $\bar{\tau}_{c}$ such
that the vectors 
\[
AJ(\bar{\tau}_{c})\psi,\frac{d}{d\tau_{1}}J(\bar{\tau}_{c})\psi,\ldots,\frac{d}{d\tau_{k-1}}J(\bar{\tau}_{c})\psi
\]
are linearly independent. By lemmas \ref{InHyperPlane} and \ref{independendLoc}
there exists $\tau_{k}\geqslant0$ such that the vectors 
\[
AJ(\bar{\tau}_{c})\psi,\frac{d}{d\tau_{1}}J(\bar{\tau}_{c})\psi,\ldots,\frac{d}{d\tau_{k-1}}J(\bar{\tau}_{c})\psi,Ae^{-\tau_{k}A}g_{1},e^{-\tau_{k}A}g_{1}
\]
are linearly independent and $a_{0}(\bar{\tau})=-2(e^{\tau_{k}A}J(\bar{\tau}_{c})\psi,g_{1})_{2}\ne0$.
Thus $\aleph=k+1$ and induction step is proved. $\blacktriangledown$

\begin{lemma} \label{InHyperPlane} For any vector $\psi\in L\setminus\{0\}$
the set 
\[
\{t>0:\ (e^{tA}\psi,g_{1})_{2}\ne0\}\subset\mathbb{R}_{+}
\]
is open end everywhere dense subset. \end{lemma} Proof. To prove
this note the following equalities 
\[
(e^{tA}\psi,g_{1})_{2}=(e^{tA}\psi,g_{1})_{H}=(\psi,e^{-tA}g_{1})_{H}.
\]
The second because $A^{\star}=-A$, where $A^{\star}$ be the adjoint
operator to $A$ for the scalar product $(,)_{H}$, that follows from
the equality 
\[
(Au,w)_{H}=(Vp,q')_{2}-(Vq,p')_{2}=-((Vq,p')_{2}-(p,Vq')_{2})=-(u,Aw)_{H}.
\]
for $u=(q,p)^{T},\ w=(q',p')^{T}\in L$. Then we use 
\[
(\psi,e^{-tA}g_{1})_{H}=\sum_{k=0}^{\infty}(-1)^{k}\frac{t^{k}}{k!}(\psi,A^{k}g_{1})_{H}
\]
as, by Lemma \ref{dissipSubSpaceDiscr}, the linear span of vectors
vectors $A^{k}g_{1},k=0,1,2,...$ coincides with $L$. Then by analiticity
of the left-hand part of the latter formula we get the proof. $\blacktriangledown$
\begin{lemma} \label{independendLoc} For any linearly independent
vectors $w_{1},\ldots,w_{k}\in L,\ k<2N-1$ the set 
\[
T(w_{1},\ldots,w_{k})=\{t>0:\dim\langle e^{-tA}g_{1},Ae^{-tA}g_{1},w_{1},\ldots,w_{k}\rangle=k+2\}\subset\mathbb{R}_{+}
\]
contains an open and everywhere dense subset. \end{lemma}

Proof. If $k<2N-2$, then choose vectors $w_{k+1},\ldots,w_{2N-2}$
so that $\dim\langle w_{1},\ldots,w_{2N-2}\rangle=2N-2$. Note that
\[
T(w_{1},\ldots,w_{2N-2})\subset T(w_{1},\ldots,w_{k})
\]
That is why it is sufficient to prove the assertion of the lemma for
the case $k=2N-2$. Thus firther on we assume that $k=2N-2$.

Consider the following real function on $L$: 
\[
W(u)=\det[u,Au,w_{1},\ldots,w_{2N-2}].
\]
Denote $X=\{u\in L:\ W(u)=0\}$ the set of zeros of the function $W$.
It is clear that 
\[
T(w_{1},\ldots,w_{2N-2})=\{t>0:W(e^{-tA}g_{1})\ne0\}=\{t>0:\ e^{-tA}g_{1}\notin X\}
\]

It is clear that $W$ is a quadratic function. Thus $X$ is a quadric.
Further we define the canonical type of the quadric X. Denote 
\[
L_{1}=\langle w{}_{1},\ldots,w_{2N-2}\rangle.
\]
Consider also the orthogonal complement $L_{1}^{\perp}$ to $L_{1}$
(in the scalar product $(,)_{H}$). By definition of $W$, we have
$L_{1}\subset X$. Thus in some coordinates $(u_{1},u_{2},\ldots,u_{2N})$
on $L$ the function $W$ will look like 
\[
W(u)=a_{1}u_{1}^{2}+a_{2}u_{2}^{2},
\]
for some $a_{1},a_{2}\in\mathbb{R}$ (as the basis for such coordinates
one could choose the vectors $w_{1},\ldots,w_{2N-2},u_{1},u_{2}$,
where $u_{1},u_{2}$ are the approriate coordinates of the orthogonal
compliment $L_{1}^{\perp}$. As $L_{1}\subset X$, then the form will
not depend on the first $2N-2$ coordinates, and one can choose two
coordinates in $L_{1}^{\perp}$ by method of Lagrange).

If we show that there is exist a vector $u$ such that $W(u)\ne0$,
than it will follows that $a_{1}$ and $a_{2}$ cannot be simultaneously
zero. And consequently the canonical type of $X$ should be one of
three types: a $2N-2$-dimensional hyperplane, $2N-1$-dimensional
hyperplane, the union of two $2N-1$ hyperplanes. Then applying the
same argument as in lemma \ref{InHyperPlane} we get the proof of
lemma \ref{independendLoc}.

Let show that that there is exist a vector $u$ such that $W(u)\ne0$.

For vector $u\in L$ denote $u^{\perp},(Au)^{\perp}$ the orthogonal
projections on $L_{1}^{\perp}$ of the vectors $u,Au$ correspondingly.
As the determinant is polilinear we have 
\[
W(u)=\det[u^{\perp},(Au)^{\perp},w_{1},\ldots,w_{2N-2}].
\]
That is why $X$ is the set of all $u\in L$ such that $u^{\perp}$and
$(Au)^{\perp}$ are linearly dependent. Note that for any $u=(q,p)^{T}\in L$
\[
(u,Au)_{H}=(Vq,p)_{2}-(p,Vq)_{2}=0.
\]
Consider two cases: 
\begin{enumerate}
\item Subspace $L_{1}$ is invariant with respect to $A$. Let us show then
$L_{1}^{\perp}$ is also invariant with respect to $A$. In fact,
for $u\in L_{1}^{\perp},\ v\in L_{1}$ we have 
\[
(Au,v)_{H}=(u,A^{\star}v)_{H}=-(u,Av)_{H}=0,
\]
as $Av\in L_{1}$ by invariance of $L_{1}$. Then $L_{1}^{\perp}$
is invariant with respect to $A$ and $X\cap L_{1}^{\perp}$ is the
set of all $u\in L_{1}^{\perp}$ such that $u$ and $Au$ are linearly
dependent. As the spectrum of $A$ is pure imaginary and does not
contain zero, then for all $u\ne0\in L_{1}^{\perp}$ the vectors $u$
and $Au$ are linearly independent. Then 
\[
X\cap L_{1}^{\perp}=\{0\}
\]
and the proof is finished in this case. 
\item Subspace $L_{1}$ is not invariant with respect to $A$. In this case
there exists vector $v\in L_{1}$ such that $u=(Av)^{\perp}\ne0$.
Consider vector $\tilde{u}\in L_{1}^{\perp}$ such that $\tilde{u},u$
are linearly independent and $\tilde{u}',u$ are linearly dependent,
where $\tilde{u}'\in L_{1}^{\perp}$ is a vector orthogonal to $\tilde{u}$
in $L_{1}^{\perp}$, again with respect to scalar product $(,)_{H}$.
As $(\tilde{u},A\tilde{u})_{H}=0$, then for the projection $(A\tilde{u})^{\perp}$
of the vector $A\tilde{u}$ on the subspace $L_{1}^{\perp}$ we have
\[
(A\tilde{u})^{\perp}=c\tilde{u}',
\]
for some $c\in\mathbb{R}$. If $c\ne0$, then it is clear that $W(\tilde{u})\ne0$.
Assume that $c=0$ and consider the vector 
\[
u^{*}=v+\tilde{u}.
\]
Then 
\[
(u^{*})^{\perp}=\tilde{u},\quad(Au^{*})^{\perp}=(Av+A\tilde{u})^{\perp}=u+(A\tilde{u})^{\perp}=u.
\]
By definition of $\tilde{u}$ the vectors $\tilde{u}$ and $u$ are
linearly independent. Then $(u^{*})^{\perp}$ and $(Au^{*})^{\perp}$
are also linearly independent, and then $W(u^{*})\ne0$. The proof
in this case is also finished $\blacktriangledown$ 
\end{enumerate}

\subsection{Proof of theorem 1}

Fix two points $\psi_{1},\psi_{2}\in\mathcal{M}$. We want to show
that for some $m^{*}$, not depending on $\psi_{1},\psi_{2}$, there
exist $\tau_{1},\ldots,\tau_{m^{*}}\geqslant0$ such that 
\[
J(\tau_{1},\ldots,\tau_{m^{*}})\psi_{1}=\psi_{2}
\]

By local covering theorem there exists point $\bar{\tau}=(\tau_{1},\ldots,\tau_{m})\in\Omega_{m}$
such that there exist neighborhood $O(\bar{\tau})\subset\Omega_{m}$
of $\bar{\tau}$ and neighborhood $O(\psi_{1}^{*})\subset\mathcal{M}$
of $\psi_{1}^{*}=J^{\psi_{1}}(\bar{\tau})=J(\tau_{1},\ldots,\tau_{m})\psi_{1}$
such that 
\begin{equation}
J^{\psi_{1}}(O(\tau))=O(\psi_{1}^{*}).\label{L060120143}
\end{equation}
Consider $\epsilon$-neighborhood $O_{\epsilon}(\psi_{1}^{*})$ of
$\psi_{1}^{*}$ in the norm, corresponding to the scalar product $(,)_{H}$,
such that 
\[
O_{\epsilon}(\psi_{1}^{*})=\{\psi\in\mathcal{M}:\ ||\psi-\psi_{1}^{*}||_{H}<\epsilon\}\subset O(\psi_{1}^{*})
\]
By closure theorem there exist $\tau'_{1},\ldots,\tau'_{m}\geqslant0$
such that 
\begin{equation}
||J(\tau'_{1},\ldots,\tau'_{m})\psi_{1}^{*}-\psi_{2}||_{H}<\epsilon.\label{L060120142}
\end{equation}
Note that for fixed $\tau'_{1},\ldots,\tau'_{m}$ the map $J(\tau'_{1},\ldots,\tau'_{m}):L\to L$
is non-degenerate and conserves the norm $||\ ||_{H}$. That is why
from (\ref{L060120142}) the following inequality follows 
\[
||\psi_{1}^{*}-J^{-1}(\tau'_{1},\ldots,\tau'_{m})\psi_{2}||_{H}<\epsilon.
\]
Thus, $J^{-1}(\tau'_{1},\ldots,\tau'_{m})\psi_{2}\in O_{\epsilon}(\psi_{1}^{*})$
and by (\ref{L060120143}) there is $\bar{\tau}^{*}=(\tau_{1}^{*},\ldots,\tau_{m}^{*})\in O(\bar{\tau})$
such that 
\[
J(\tau_{1}^{*},\ldots,\tau_{m}^{*})\psi_{1}=J^{-1}(\tau'_{1},\ldots,\tau'_{m})\psi_{2}.
\]
This can be rewritten as 
\[
J(\tau_{1}^{*},\ldots,\tau_{m}^{*},\tau'_{1},\ldots,\tau'_{m})\psi_{1}=J(\tau'_{1},\ldots,\tau'_{m})J(\tau_{1}^{*},\ldots,\tau_{m}^{*})\psi_{1}=\psi_{2}.
\]
$\blacktriangledown$

\pagebreak

\section{Stochastic part - proof of convergence theorem }

\subsection{Embedded process}

Here we consider the sequence $\psi_{k}=\psi(t_{k})$, which is a
discrete time Markov chain (embedded chain) with state space $\mathcal{M}$.
We prove that it is (Markov) ergodic as it is defined in the following
theorem concerning more general class of discrete time Markov chains.
Namely, we consider Markov chains $\xi_{n}$ on compact state space
$X$ with Borel $\sigma$-algebra $\mathcal{B}(X)$ (with countable
basis) and transition probability kernels $P(x,A)$, which are probability
measures on $\mathcal{B}(X)$ for any $x\in X$ and measurable functions
on $X$ for any $A\subset X$. We will consider the class of such
chains satisfying the following assumptions:

A1) for some integer $m\geq1$ and any $x\in X$ the $m$-step transition
probability $P^{m}(x,.)$ is equivalent to some finite non-negative
measure $\mu$ such that $\mu(O)>0$ for any open set $O\subset X$.
Moreover, for any $x$ there exists $m$-step transition density $p^{m}(x,y)$
(with respect to $\mu$), which is measurable on $\mathcal{M}\times\mathcal{M}$;

A2) for any open $O\subset X$ the function $P(x,O)$ is lower semi-continuous.

\begin{theorem}\label{Markov_chain}

Under assumptions A1, A2 the Markov chain $\xi_{n}$ is ergodic, that
is there exists probability measure $\pi$ on $X$ such that 
\begin{equation}
\sup_{A\in\mathcal{B}(X)}|P^{n}(x,A)-\pi(A)|\rightarrow0\label{ergodicity}
\end{equation}
as $n\to\infty$, uniformly in $x$.

\end{theorem}

This theorem follows probably from the existing deep theory of such
Markov chains with continuous state space, see for example \cite{Orey,MT},
but we did not find the statement we need, and for the reader's convenience
we give a short proof below, using the ideas from \cite{Orey,MT}.

\begin{corollary}The embedded chain $\psi_{k}$ has the unique invariant
measure $\pi$, and moreover it is ergodic as in theorem \ref{Markov_chain}. 

\end{corollary}

We have only to prove the properties A1) and A2) for our embedded
chain. For the rest of this section the integer $m$ is the same as
in the covering theorem.

\paragraph{Proof of A1)}

We will prove that the measures $\pi$ and $P^{m}(\psi,A)$ are equivalent
for any $\psi$.

\begin{lemma} \label{zeroSet-1} For any measurable $B\subset\mathcal{M}$
its Liouville measure $\pi(B)=0$ iff the Lebesgue measure $\lambda$
of the set $(J_{m}^{\psi})^{-1}(B)$ in $\Omega_{m}$ is zero. \end{lemma}

Proof. 1) Assume that for some $B\subset\mathcal{M}$ we have $\pi(B)=0$.
Let us show that $\lambda((J_{m}^{\psi})^{-1}(B))=0$. Let $A_{cr}$
be the set of critical points of the map $J_{m}^{\psi}$ (that is
points $\overline{\tau}=(\tau_{1},\ldots,\tau_{m})$ where the rank
is not maximal) and let $E=J_{m}^{\psi}(A_{cr})\subset\mathcal{M}$
be the set of critical values of $J_{m}^{\psi}$. By Sard's theorem
$\pi(E)=0$. But as $J_{m}^{\psi}(\Omega^{\psi})=\mathcal{M}$, then
there exists non-critical point $\overline{\tau}=(\tau_{1},\ldots,\tau_{m})\in\Omega_{m}$,
that is such that the rank of $dJ_{m}^{\psi}$ at this point equals
$2N-1$. As the map $J_{m}^{\psi}$ is analytic in the variables $\tau_{1},\ldots,\tau_{m}$,
the set of points $A_{cr}$, where the rank is less than $2N-1$,
has Lebesgue measure zero. Then the equality $\lambda((J_{m}^{\psi})^{-1}(B))=0$
follows from theorem 1 of \cite{Ponomarev}.

2) Assume that for some $B\subset\mathcal{M}$ we have $\pi(B)>0$,
and let us show that $\lambda((J_{m}^{\psi})^{-1}(B))>0$. By Lebesgue
differentiation theorem there exists point $\psi'\in\mathcal{M}\setminus E$
and its neighbourhood $O(\psi')$ such that $\pi(O(\psi')\cap B)>0$.
Then there is point $\overline{\tau}=\overline{\tau}(\psi')\in(J_{m}^{\psi})^{-1}(\psi')$
and some its neighborhood $O(\overline{\tau})\subset\Omega_{m}$,
so that the restriction of $J_{m}^{\psi}$ on $O(\overline{\tau})$
is a submersion. Then $\pi(O(\psi')\cap B)>0$ implies $\lambda((J_{m}^{\psi})^{-1}(B)\cap O(\overline{\tau}))>0$.
$\blacktriangledown$

Denote $\rho^{(m)}$ the product of $m$ densities $\rho$, then as
for any $B\subset\mathcal{M}$ 
\[
P^{m}(\psi,B)=\int_{(J_{m}^{\psi})^{-1}(B)}\rho^{(m)}(\overline{\tau})d\overline{\tau}.
\]
by lemma \ref{zeroSet-1} we get that $P^{m}$ and $\pi$ are equivalent
measures.

The proof of measurability of the transition density there is in theorem
1, p. 180 of \cite{Skorohod}, and in Proposition 1.1, p.5, of \cite{Orey}.

\paragraph{Proof of A2)}

\begin{lemma} Our Markov chain $\psi_{k}$ is a weak Feller chain,
that is for any open $O\subset\mathcal{M}$ the transition probability
$P(\psi,O)$ is lower semicontinuous in $\psi$. \end{lemma}

Proof. For any $\psi$ denote $\mathbf{1}_{\psi}(\tau)$ the indicator
function on $R_{+}$, that is $\mathbf{1}_{\psi}(\tau)=1$ if $Ie^{\tau A}\psi\in O$,
and zero otherwise. Then we have 
\[
P(\psi,O)=\int_{\mathbb{R}_{+}}\mathbf{1}_{\psi}(\tau)\rho(\tau)d\tau,
\]
Let $\psi_{n}\rightarrow\psi,\psi_{n}\in\mathcal{M}.$ as $n\rightarrow\infty$.
Fix $\tau\geqslant0$ and consider two cases:

1. $Ie^{\tau A}\psi\in O$, then starting from some $n$ the inclusion
$Ie^{\tau A}\psi_{n}\in O$ holds, as $O$ is open. That is why 
\[
\lim_{n\rightarrow\infty}\mathbf{1}_{\psi_{n}}(\tau)=\mathbf{1}_{\psi}(\tau)=1
\]
2. $Ie^{\tau A}\psi\notin O$. Then 
\[
\liminf_{n}\mathbf{1}_{\psi_{n}}(\tau)\geqslant\mathbf{1}_{\psi}(\tau)=0.
\]
Thus for any $\tau$ 
\[
\liminf_{n}\mathbf{1}_{\psi_{n}}(\tau)\geqslant\mathbf{1}_{\psi}(\tau)
\]
Then by Fatou lemma 
\[
\liminf_{n}P(\psi_{n},O)\geqslant P(\psi,O).
\]
$\blacktriangledown$

\subsection{Proof of theorem \ref{Markov_chain}}

\paragraph{Small sets}

We will need the following important definition.

\begin{definition}

Below $\nu$ will be any non-zero non-negative measure not necessarily
probabilistic. The Borel subset $C\subset X$ is called $(\nu,n)$-small
(or simply small, if it is $(\nu,n)$-small for some integer $n>0$
and some $\nu$ ) if for any $x\in C$ and any Borel set $B$ 
\[
P^{n}(x,B)\geq\nu(B)
\]

\end{definition}

\begin{lemma} \label{minorSetExist} Assume $A1).$ Then for some
$n_{1}\geqslant1$ and some $\nu$ there exists $(\nu,n_{1})$-small
subset $C\in\mathcal{B}(X)$ such that $\nu(C)>0$. \end{lemma}

Proof. It follows from the assumption A1) that for any $n\geqslant m$
there is measurable function $p_{n}(x,y)$ such that 
\[
P^{n}(x,B)=\int_{B}p_{n}(x,y)\mu(dy),
\]
for any $x\in X,\ B\in\mathcal{B}(X)$. Moreover, $p_{n}(x,y)>0$
for almost any $(x,y)\in X\times X$ (with respect to $\mu\times\mu$).

We will prove more. Namely, that for some $n\geqslant m$ there exist
sets $B_{1},B_{2}\in\mathcal{B}(X)$ of positive measure $\mu$ and
some constant $\delta>0$ such that for all $x\in B_{1},\ y\in B_{2}$
we have 
\begin{equation}
p_{n}(x,y)>\delta\label{density_delta}
\end{equation}
As the density $p_{m}(x,y)$ is measurable and almost everywhere positive,
one can find number $c>0$ so that the sets 
\begin{align*}
 & A=\{(x,y)\in X\times X:p_{m}(x,y)>c\}\\
 & \{(x,y,z)\in X\times X\times X:\ (x,y)\in A\ \mbox{\textcyr{\char232}}\ (y,z)\in A\}=(A\times X)\cap(X\times A)
\end{align*}
have positive measures $\mu^{2}=\mu\times\mu$ on $X\times X$ and
$\mu^{3}=\mu\times\mu\times\mu$ on $X\times X\times X$ correspondingly.
Denote $O_{r}(x)\subset X$ the open neighborhood of $x\in X$ of
radius $r$, and put $O_{r}(x,y)=O_{r}(x)\times O_{r}(y)\subset X\times X$.
By Lebesgue differentiation theorem there exists a set $A_{0}$ of
zero measure $\mu\times\mu$ such that for any $(x,y)\in A\setminus A_{0}$
\[
\lim_{r\rightarrow0+}\frac{\mu^{2}(A\cap O_{r}(x,y))}{\mu^{2}(O_{r}(x,y))}=1.
\]
It follows, that the set $((A\setminus A_{0})\times X)\cap(X\times(A\setminus A_{0}))$
also has positive measure. Consider some point $(x^{*},y^{*},z^{*})$
in this set. Choose $r$ so that the following inequalities hold:
\begin{align*}
\mu^{2}(A\cap O_{r}(x^{*},y^{*}))> & \ \frac{3}{4}\mu^{2}(O_{r}(x^{*},y^{*}))\\
\mu^{2}(A\cap O_{r}(y^{*},z^{*}))> & \ \frac{3}{4}\mu^{2}(O_{r}(y^{*},z^{*}))
\end{align*}
For any $x\in X$ put 
\[
A_{L}(x)=\{y\in X:(x,y)\in A\},\quad A_{R}(z)=\{y\in X:(y,z)\in A\}
\]
Define $B_{1}$ by 
\[
B_{1}=\{x\in O_{r}(x^{*}):\mu(A_{L}(x)\cap O_{r}(y^{*}))>\frac{3}{4}\mu(O_{r}(y^{*}))\}
\]
Otherwise speaking, the set $B_{1}$ consists of the points $x\in O_{r}(x^{*})$,
for which the set $A_{L}(x)$ is sufficiently large inside $O_{r}(y^{*})$.
Let us show that $B_{1}$ has positive measure. Assume the contrary
- that for almost all points of $O_{r}(x^{*})$ the inequality $\mu(A_{L}(x)\cap O_{r}(y^{*}))\leqslant\frac{3}{4}\mu(O_{r}(y^{*}))$
holds. Then by Fubini theorem 
\[
\mu^{2}(A\cap O_{r}(x^{*},y^{*}))=\int_{O_{r}(x^{*})}\mu(A_{L}(x)\cap O_{r}(y^{*}))\mu(dx)\leqslant\frac{3}{4}\mu(O_{r}(y^{*}))\mu(O_{r}(x^{*}))=\frac{3}{4}(\mu\times\mu)(O_{r}(x^{*},y^{*})).
\]
That contradicts the choice of the points $x^{*},y^{*}$. Thus, $\mu(B_{1})>0$.
Similarly, one can show that the set 
\[
B_{2}=\{z\in O_{r}(z^{*}):\mu(A_{R}(z)\cap O_{r}(y^{*}))>\frac{3}{4}\mu(O_{r}(y^{*}))\}
\]
has positive measure $\mu$. But from the definition of the sets $B_{1},\ B_{2}$
it follows that for any points $x\in B_{1},\ z\in B_{2}$ the following
inequality holds 
\[
\mu(A_{L}(x)\cap A_{R}(z))\geqslant\frac{1}{2}\mu(O_{r}(y^{*})).
\]
Also we have the estimates for the density $p_{2m}$ and any $x\in B_{1},\ z\in B_{2}$
\begin{align*}
 & p_{2m}(x,z)=\int_{X}p_{m}(x,y)p_{m}(y,z)\mu(dy)\geqslant\int_{A_{L}(x)\cap A_{R}(z)}p_{m}(x,y)p_{m}(y,z)\mu(dy)>\\
 & >c^{2}\mu(A_{L}(x)\cap A_{R}(z))\geq\frac{c^{2}}{2}\mu(O_{r}(y^{*})).
\end{align*}

Thus we have proved (\ref{density_delta}). Now we finish the proof
of lemma \ref{minorSetExist}. Note that for any $x\in B_{1}$ and
any $B\in\mathcal{B}(X)$ we have 
\[
P^{n}(x,B)=\int_{B}p_{n}(x,y)\mu(dy)\geqslant\int_{B\cap B_{2}}p_{n}(x,y)\mu(dy)\geqslant\delta\mu(B\cap B_{2}).
\]
Moreover, as the sets $B_{1},B_{2}$ have positive measure, then there
exists subset $C\subset B_{2}$ of positive measure $\mu$ and constant
$\delta'>0$ such that $P^{m}(x,B_{1})>\delta'$ for all $x\in C$.
It follows that for all $x\in C,\ B\in\mathcal{B}(X)$ the following
inequalities hold: 
\[
P^{m+n}(x,B)=\int_{X}P^{m}(x,dy)P^{n}(y,B)\geqslant\int_{B_{1}}P^{m}(x,dy)P^{n}(y,B)\geqslant\delta'\delta\mu(B\cap B_{2}).
\]
Thus, $C$ is $(\nu,n+m)$-small, where $\nu(B)=\delta\delta'\mu(B\cap B_{2})$,
and moreover $\nu(C)>0$. $\blacktriangledown$

\begin{lemma} \label{minorConst} Let $C\in\mathcal{B}(X)$ be $(\nu,n_{1})$-small
and assume that for any $x$ from some set $D\in\mathcal{B}(X)$ 
\[
P^{n}(x,C)>\delta.
\]
for some $\delta>0$ and $n\geqslant1$. Then the set $D$ is $(\delta\nu,n+n_{1})$-small.
\end{lemma}

Proof. For any $x\in D$ and any $B\in\mathcal{B}(X)$ by semigroup
property 
\[
P^{n+n_{1}}(x,B)=\int_{X}P^{n}(x,dy)P^{n_{1}}(y,B)\geqslant\int_{C}P^{n}(x,dy)P^{n_{1}}(y,B)\geqslant P^{n}(x,C)\nu(B)\geqslant\delta\nu(B),
\]
$\blacktriangledown$

\begin{lemma}\label{minorCond}

Under assumptions A1) and A2) the set $X$ itself is small.

\end{lemma}

Proof. Consider some $(\nu,n_{1})$-small subset $C\in\mathcal{B}(X)$
of lemma \ref{minorSetExist}. For $k=1,2,\ldots$ introduce the subsets
\[
A_{k}=\{x\in X:\ P^{m}(x,C)>\frac{1}{k}\},
\]
where $m$ is defined in A1). From lemma \ref{minorConst} we have
that $A_{k}$ is a $(\frac{1}{k}\nu,m+n_{1})$-small set. Let us prove
that its closure is also a small set. Note that the measure $\nu$
is regular, that is for any Borel set $B$ 
\[
\nu(B)=\sup\{\nu(K)\},
\]
where the supremum is over all compact sets $K\subset B$. As $\nu(C)>0$,
there exists compact set $K\subset C$ such that for all $x\in A_{k}$
\[
P^{m+n_{1}}(x,K)\geqslant\frac{1}{k}\nu(K)=\delta>0.
\]
Let $\{x_{n}\}_{n=1,2\ldots}\in A_{k}$ and $x_{n}\rightarrow x$,
then by semi-continuity of the transition probability we get: 
\[
P^{m+n_{1}}(x,K)=1-P^{m+n_{1}}(x,X\setminus K)\geqslant1-\lim\inf P^{m+n_{1}}(x_{n},X\setminus K)=\limsup P^{m+n_{1}}(x_{n},K)\geqslant\delta.
\]
Thus, for any $x\in\bar{A}_{k}$ we have $P^{m+n_{1}}(x,K)>\delta/2$,
and applying lemma \ref{minorConst}, we get that $\bar{A}_{k}$ is
a $(\frac{\delta}{2}\nu,n_{2})$-small set for some $n_{2}$.

But also by assumption A1 
\[
X=\cup_{k=1}^{\infty}A_{k}
\]
As $X$ is not a countable union of sets which are nowhere dense,
then for some $k$ there is an open subset $O\subset\bar{A}_{k}$.
Thus for any $x\in X$ and any $B\in\mathcal{B}(X)$, we have 
\begin{equation}
P^{m+n_{2}}(x,B)=\int_{X}P^{m}(x,dy)P^{n_{2}}(y,A)\geqslant\int_{O}P^{m}(x,dy)P^{n_{2}}(y,B)\geqslant\frac{\delta}{2}\nu(B)P^{m}(x,O).\label{mneq}
\end{equation}
As the measures $P(x,\cdot)$ and $\mu$ are equivalent for any $x\in X$,
we have $P^{m}(x,O)>0$. But $P^{m}(x,O)$ is lower semi-continuous,
and thus attains minimum on the compact $X$. It follows that 
\[
P^{m}(x,O)>\delta'
\]
for some $\delta'>0$ and any $x\in X$. Using inequality (\ref{mneq}),
we get the proof.

\paragraph{Proof of theorem \ref{Markov_chain}}

For$A\in\mathcal{B}(X)$ and $n\geqslant1$ denote 
\[
I_{n}(A)=\inf_{x\in X}P^{n}(x,A),\quad S_{n}(A)=\sup_{x\in X}P^{n}(x,A).
\]
Note that 
\[
I_{n+1}(A)=\inf_{x\in X}\int_{X}P(x,dy)P^{n}(y,A)\geqslant\inf_{x\in X}\int_{X}P(x,dy)I_{n}(A)=I_{n}(A).
\]
Thus,, for fixed $A\in\mathcal{B}(X)$ the sequence $I_{n}(A)$ is
non-decreasing. Similarly the sequence $S_{n}(A)$ is non-increasing.
We shall prove that $S_{n}(A)-I_{n}(A)$ tends to zero as $n\rightarrow\infty$.

Take number $N$ and measure $\nu$ as in Lemma \ref{minorCond}.
Then for any $n\geqslant1$ 
\[
P^{n+N}(x,A)=\int_{X}P^{N}(x,dy)P^{n}(y,A)=\int_{X}(P^{N}(x,dy)-\nu(dy))P^{n}(y,A)+\int_{X}\nu(dy)P^{n}(y,A)
\]
As the measure $P^{N}(x,\cdot)-\nu(\cdot)$ is non-negative, we get
from this equality that 
\[
P^{n+N}(x,A)\geqslant I_{n}(A)\int_{X}(P^{N}(x,dy)-\nu(dy))+\int_{X}\nu(dy)P^{n}(y,A)=(1-\nu(X))I_{n}(A)+\int_{X}\nu(dy)P^{n}(y,A)
\]
and then 
\[
I_{n+N}(A)\geqslant(1-\delta)I_{n}(A)+c_{1},\quad\delta=\nu(X),\ c_{1}=\int_{X}\nu(dy)P^{n}(y,A).
\]
Similarly we get the upper bound for $S_{n+N}(A)$: 
\[
P^{n+N}(x,A)\leq(1-\nu(X))S_{n}(A)+\int\nu(dy)P^{n}(y,A)\leq(1-\delta)S_{n}+c_{1}
\]
and 
\[
S_{n+N}(A)\leqslant(1-\delta)S_{n}(A)+c_{1}.
\]
Then the difference is estimated as follows 
\[
S_{n+N}(A)-I_{n+N}(A)\leqslant(1-\delta)(S_{n}(A)-I_{n}(A)).
\]
From the last inequality and monotonicity of the corresponding sequences
it follows that for any $x\in X$ there exist the following limits
and that they are equal 
\[
\lim_{n\rightarrow\infty}I_{n}(A)=\lim_{n\rightarrow\infty}S_{n}(A)=\lim_{n\rightarrow\infty}P_{n}(x,A)=\pi(A),
\]
and moreover the convergence is uniform in $A\in\mathcal{B}(X)$ and
in $x$. $\blacktriangledown$

\subsection{Proof of theorem \ref{th_erg}}

We will use the following theorem (strong law of large numbers) for
discrete time Markov chains on arbitrary state space $X$ equipped
with $\sigma$-algebra $\mathcal{A}$. Let $P^{n}(x,B)$ be $n$-step
transition probability assumed to be measurable on $X$ for any $B\in\mathcal{A}$
and is a probability measure on $(X,\mathcal{A})$ for any $x$. Let
us assume that there exists invariant measure $\pi$ on $(X,\mathcal{A})$
such that uniformly in $x$
\[
\sup_{A\in\mathcal{A}}|P^{n}(x,A)-\pi(A)|\rightarrow0,\quad n\rightarrow\infty
\]

Denote $P_{x}$ the measure on trajectories $(x_{0}=x,x_{1},x_{2},...)$
with initial point $x$. Under these conditions the following assertion
holds.

\begin{theorem} \label{strongLaw-1} For any $f\in L^{1}(X,\pi)$
and any $x\in X$ we have $P_{x}$-a.s. 
\[
\lim_{n\rightarrow\infty}\frac{1}{n}\sum_{k=0}^{n}f(x_{k})=\int_{X}f(x)\pi(dx)
\]
 \end{theorem}

Proof. See \cite{Revuz}, p. 140, and \cite{Skorohod}, p. 209. 

To prove theorem \ref{th_erg} we need the following lemma.

\begin{lemma} For any measurable bounded function $f$ on $\mathcal{M}$
and any initial state $\psi(0)=\psi$ the following limit holds a.s.
\[
\lim_{N\rightarrow\infty}M_{f}(t_{N})=\pi(f)
\]
\end{lemma} 

Proof. Denote $X_{k}=(\psi_{k},\tau_{k+1}),\ k=0,1,\ldots$ the Markov
chain with values in $X=\mathcal{M}\times\mathbb{R}_{+}$. Then

\begin{equation}
\int_{t_{k}}^{t_{k+1}}f(\psi(s))ds=\int_{t_{k}}^{t_{k+1}}f(e^{(s-t_{k})A}\psi_{k})ds=\int_{0}^{\tau_{k+1}}f(e^{sA}\psi_{k})ds=F(X_{k})\label{tktkp1-1}
\end{equation}
 where 
\[
F(\psi,t)=\int_{0}^{t}f(e^{sA}\psi)ds,\ (\psi,t)\in X.
\]
Then 
\begin{equation}
M_{f}(t_{N})=\frac{1}{t_{N}}\sum_{k=0}^{N-1}\int_{t_{k}}^{t_{k+1}}f(\psi(s))ds=\frac{1}{t_{N}}\sum_{k=0}^{N-1}F(X_{k})\label{MasSum-1}
\end{equation}

It is easy to show that $X_{k}$ has invariant measure $\mu=\pi\times P_{\tau},P_{\tau}=\rho ds,$
satisfies the conditions of theorem \ref{strongLaw-1} as $\psi_{k}$
satisfies it. Then 
\[
\lim_{N\rightarrow\infty}\frac{1}{N}\sum_{k=0}^{N-1}F(X_{k})=\mu(F)=\int_{X}F(\psi,s)d\mu
\]
where
\[
\mu(F)=\int_{\mathbb{R}_{+}}P_{\tau}(dt)\int_{\mathcal{M}}\pi(d\psi)\int_{0}^{t}dsf(e^{sA}\psi)=\int_{\mathbb{R}_{+}}P_{\tau}(dt)\int_{0}^{t}ds\int_{\mathcal{M}}\pi(d\psi)f(e^{sA}\psi)=
\]
\[
=\pi(f)\int_{\mathbb{R}_{+}}P_{\tau}(dt)\int_{0}^{t}ds=\pi(f)E\tau_{1}
\]
Moreover, by strong law of large numbers for independent random varoables
$\tau_{k}$ we have 
\[
\lim_{N\rightarrow\infty}\frac{t_{N}}{N}=E\tau_{1}
\]
Then by (\ref{MasSum-1}) we get the proof of the lemma.$\blacktriangledown$

To prove theorem \ref{th_erg} we have to estimate the difference
between $M_{f}(t)$ and $M_{f}(t_{N})$. Using the boundedness $|f(\psi)|\leqslant c$
we have 
\begin{equation}
|M_{f}(t)-M_{f}(t_{N})|\leqslant|\frac{1}{t}\int_{t_{N}}^{t}f(\psi(s))ds|+\frac{|t-t_{N}|}{t}|M_{f}(t_{N})|\leqslant\frac{|t-t_{N}|}{t}(c+|M_{f}(t_{N})|).\label{diffEstim-1}
\end{equation}
For any $t>0$ define the random index $N(t)$ so that 
\[
t_{N(t)}\leqslant t<t_{N(t)+1}.
\]
and note that 
\[
\frac{|t-t_{N(t)}|}{t}\leqslant\frac{\tau_{N(t)+1}}{t_{N(t)}}=\frac{\tau_{N(t)+1}}{\sum_{k=1}^{N(t)}\tau_{k}}.
\]
As $E\tau_{1}<\infty$, the law of large numbers, as $N\to\infty$,
gives a.s. 
\[
\frac{\tau_{N+1}}{\sum_{k=1}^{N}\tau_{k}}\rightarrow0
\]
But $N(t)\rightarrow\infty$ as $t\rightarrow\infty$. Then the right-hand
side of (\ref{diffEstim-1}) tends to $0$ a.s. as $N=N(t)$ and $t\rightarrow\infty$.
$\blacktriangledown$

\pagebreak

\section{Appendix}

\subsection{Proof of lemma \ref{L:action-angle}}

For any $t\geqslant0$ one can show that 
\[
e^{tA}=\left(\begin{array}{cc}
\cos(t\sqrt{V}) & (\sqrt{V})^{-1}\sin(t\sqrt{V})\\
-\sqrt{V}\sin(t\sqrt{V}) & \cos(t\sqrt{V})
\end{array}\right),
\]
where $\sqrt{V}$ is the positive square root of the matrix $V$.
Then for any $k=1,\ldots,N$ and $t\geqslant0$. 
\begin{align}
e^{tA}Q_{k}= & \ \cos(\omega_{k}t)Q_{k}-\omega_{k}\sin(\omega_{k}t)P_{k},\label{expQ}\\
e^{tA}P_{k}= & \ \frac{\sin(\omega_{k}t)}{\omega_{k}}Q_{k}+\cos(\omega_{k}t)P_{k},\label{expP}
\end{align}
Using (\ref{expQ})-(\ref{expP}) we have for any $\psi$ 
\[
\psi(t)=e^{tA}\psi=\sum_{k=1}^{N}\left(\cos(\omega_{k}t)\tilde{q}_{k}+\frac{\sin(\omega_{k}t)}{\omega_{k}}\tilde{p}_{k}\right)Q_{k}+\sum_{k=1}^{N}\left(-\omega_{k}\sin(\omega_{k}t)\tilde{q}_{k}+\cos(\omega_{k}t)\tilde{p}_{k}\right)P_{k}.
\]
and then 
\begin{align*}
\tilde{q}_{k}(t)= & \ \cos(\omega_{k}t)\tilde{q}_{k}+\frac{\sin(\omega_{k}t)}{\omega_{k}}\tilde{p}_{k}\\
\tilde{p}_{k}(t)= & \ -\omega_{k}\sin(\omega_{k}t)\tilde{q}_{k}+\cos(\omega_{k}t)\tilde{p}_{k}.
\end{align*}
From these two formulas we see that pair of functions $(\tilde{q}_{k}(t),\tilde{p}_{k}(t)),k=1,\ldots,N,$
corresponds to the dynamics of one-dimensional oscillator of unit
mass and frequency $\omega_{k}$, where $\tilde{q}_{k}(t)$ is the
oscillator coordinate and $\tilde{p}_{k}(t)$ is its momentum. Thus
the dynamics $e^{tA}\psi$ is isomorphic to the uniform movement on
the torus with velocity $(\omega_{1}^{2},\ldots,\omega_{N}^{2})$.
This gives the second assertion.

\subsection{Case $N=1$}

Note the identity for any $\psi$ 
\begin{equation}
Ie^{At_{1}}\psi=e^{-At_{1}}I\psi\label{identity_1}
\end{equation}
Let us first prove the proposition for $t=T_{0}$, $\psi=\psi_{0}=(1,0)$
and arbitrary $\psi'$. Then for any define the one-to-one mapping
$W:(0,\frac{1}{2}T_{0})\to S$. namely for any $t_{1}\in(0,\frac{1}{2}T_{0})$
we define 
\[
W(t')=\psi'=e^{A(T_{0}-t_{1})}Ie^{At_{1}}\psi_{0}=e^{A(T_{0}-2t_{1})}\psi_{0}
\]
Then it is sufficient to take $t'=\frac{T_{0}-t''}{2}$ and choose
minimal $t''>0$ so that 
\[
\psi'=e^{At''}\psi_{0}
\]
For $t>T_{0}$ the proof is quite similar but we will not need this
case to prove convergence.

\subsection{Mixing subspace}

Here we give some properties of the mixing space. The following two
lemmas show how the dimension of $L_{-}$can be explicitely characterized.

\begin{lemma} \label{dissipSubSpaceDiscr} The space $L_{-}$ is
invariant with respect to $A$. Moreover 
\[
L_{-}=\langle\{A^{k}g_{1}:\ k=0,1,\ldots\}\rangle
\]
where $\langle\ \rangle$ is the linear span of the set of vectors.
\end{lemma}

Consider also the orthogonal complement to $L_{-}$ in the scalar
product $(,)_{2}$ 
\[
L_{0}=L_{-}^{\perp}.
\]
Then it is also invariant with respect to $A$. Moreover, the vector
$\psi\in L_{0}$ iff for the hamiltonian dynamics with initial condition
$\psi=\psi(0)$ the momentum $p_{1}(t)=0$ for any $t$.

\begin{lemma} \label{0509132} Assume that the spectrum of $V$ is
simple, and let $\{v_{1},\ldots,v_{N}\}$ be the eigenvectors of $V$,
they form a basis in $R^{N}$. Then the dimension of $L_{0}$ is twice
the number of $v_{k}$ having coordinates $v_{k,1}=(e_{1},v_{k})=0$.
\end{lemma}

What occurs if condition (\ref{L201406141}) is not fulfilled, exact
formulas for the dimension of $L_{-}$ for the chain of harmonic oscillators
and for other cases see in \cite{LM_1,LM_2,LM_4}.

\subsection{Proof of lemma \ref{lemma_cube}}

We have 
\[
|\tilde{p}_{k}^{*}|=|\beta_{k}|\left|y-c\frac{1-\gamma_{k}^{2}}{4y}\right|=|\beta_{k}|\left|\frac{2\gamma_{n}^{2}+2\gamma_{n}\sqrt{\gamma_{n}^{2}+c(1-\gamma_{n}^{2})}+c(\gamma_{k}^{2}-\gamma_{n}^{2})}{4y}\right|
\]
As for any $k=1,\ldots,N$ we have $\gamma_{k}\geqslant\gamma_{n}\geqslant0$,
then the expression under module in the last formula is non-negative,
and we have 
\[
|\tilde{p}_{k}^{'}|=|\beta_{k}|\left(y-c\frac{1-\gamma_{k}^{2}}{4y}\right).
\]
Consider two cases: 
\begin{enumerate}
\item $\gamma_{k}\leqslant1.$ Show that $f^{+}(x)$ is monotone increasing,
that is its derivative 
\[
(f^{+}(x))'=\frac{1}{2}\left(1+\frac{x(1-c)}{\sqrt{x^{2}+c(1-x^{2})}}\right)>0.
\]
Thus $y=f^{+}(\gamma_{n})\leqslant f^{+}(\gamma_{k})$. Taking into
account $\gamma_{k}\leqslant1$, we have the inequalities 
\begin{align*}
 & |\tilde{p}_{k}^{'}|=|\beta_{k}|\left(y-c\frac{1-\gamma_{k}^{2}}{4y}\right)\leqslant|\beta_{k}|\left(f^{+}(\gamma_{k})-\frac{c(1-\gamma_{k}^{2})}{4f^{+}(\gamma_{k})}\right)=\\
 & =|\beta_{k}|\frac{2\gamma_{k}^{2}+2\gamma_{k}\sqrt{\gamma_{k}^{2}+c(1-\gamma_{k}^{2})}}{4f^{+}(\gamma_{k})}=|\beta_{k}|\gamma_{k}=r_{k}.
\end{align*}

\item $\gamma_{k}>1$. Then we will show that $f^{-}(\gamma_{k})\leqslant y$.
It is easy to check that $f^{-}(x)$ is increasing and then 
\[
0<f^{-}(\gamma_{k})\leqslant f^{-}(\gamma_{N}).
\]
Note that the left inequality holds because $f^{-}(1)=0$.

Let us prove that for any $x\in\mathbb{R}$ we have $f^{-}(x)<\frac{c}{2}x$.
Consider the chain of inequalities 
\[
f^{-}(x)-\frac{c}{2}x=\frac{1}{2}\left((1-c)x-\sqrt{x^{2}(1-c)+c}\right)=\frac{1}{2}\left(\frac{-c-x^{2}(1-c)c}{(1-c)x+\sqrt{x^{2}(1-c)+c}}\right)<0
\]
Thus we have proved that 
\begin{equation}
f^{-}(\gamma_{k})\leqslant f^{-}(\gamma_{N})<\frac{c}{2}\gamma_{N}.\label{eqfmin}
\end{equation}
By $f^{+}(0)=\frac{1}{2}\sqrt{c}$ and the evident inequality $(f^{+}(x))'\geqslant\frac{c}{2}$,
which holds as $c\leqslant1$, we have 
\begin{equation}
y=f^{+}(\gamma_{n})\geqslant\frac{1}{2}(c\gamma_{n}+\sqrt{c}).\label{eqy}
\end{equation}
Then by (\ref{eqfmin}), (\ref{eqy}) and $c\leqslant\frac{1}{(\gamma_{N}-\gamma_{n})^{2}}$
we get 
\[
y-f^{-}(\gamma_{k})\geqslant\frac{1}{2}(-c(\gamma_{N}-\gamma_{n})+\sqrt{c})=\frac{\sqrt{c}}{2}\left(-\sqrt{c}(\gamma_{N}-\gamma_{n})+1\right)\geqslant0.
\]
Remind that we have proved earlier that the function $f^{+}(x)$ is
increasing, then taking into account the latter inequality we have
the following bounds 
\begin{align*}
 & |\tilde{p}_{k}^{*}|=|\beta_{k}|\left(y-c\frac{1-\gamma_{k}^{2}}{4y}\right)=|\beta_{k}|\left(y+c\frac{\gamma_{k}^{2}-1}{4y}\right)\leqslant|\beta_{k}|\left(f^{+}(\gamma_{k})+c\frac{\gamma_{k}^{2}-1}{4f^{-}(\gamma_{k})}\right)=\\
 & =|\beta_{k}|\left(\frac{\gamma_{k}^{2}-\gamma_{k}^{2}-c(1-\gamma_{k}^{2})+c(\gamma_{k}^{2}-1)}{4f^{-}(\gamma_{k})}\right)=|\beta_{k}|c\frac{\gamma_{k}^{2}-1}{2f^{-}(\gamma_{k})}=|\beta_{k}|c\frac{2(\gamma_{k}^{2}-1)f^{+}(\gamma_{k})}{2c(\gamma_{k}^{2}-1)}=\\
 & =|\beta_{k}|f^{+}(\gamma_{k})\leqslant|\beta_{k}|\gamma_{k}=r_{k}.
\end{align*}

\end{enumerate}
In the last inequality we used fact that $f^{+}(x)<x$ for $x>1$
holds due to $f^{+}+f^{-}=x$. $\blacktriangledown$

\end{document}